\documentclass[twocolumn,showpacs,preprintnumbers,nofootinbib,amsmath,amssymb]{revtex4}
\usepackage{graphicx}
\usepackage{dcolumn}
\usepackage{bm}
\newcommand{\be}{\begin{equation}}
\newcommand{\ee}{\end{equation}}
\newcommand{\bef}{\begin{figure}}
\newcommand{\eef}{\end{figure}}
\newcommand{\bea}{\begin{eqnarray}}
\newcommand{\eea}{\end{eqnarray}}
\newcommand{\bei}{\begin{itemize}}
\newcommand{\eei}{\end{itemize}}
\newcommand{\bx}{{\bf x}}
\newcommand{\bk}{{\bf k}}

\newcommand{\by}{{\bf y}}

\newcommand{\bu}{{\bf u}}
\newcommand{\bR}{{\bf R}}
\newcommand{\bfeta}{{\bf \eta}}
\newcommand{\bH}{{\bf H}}
\newcommand{\buri}{{\bf u}_r^{(i)}}

\newcommand{\busi}{{\bf u}_s^{(i)}}
\newcommand{\busj}{{\bf u}_s^{(j)}}
\newcommand{\bv}{{\bf v}}
\newcommand{\bw}{{\bf w}}
\newcommand{\la}{\left<}
\newcommand{\ra}{\right>}

\begin{document}
\title{Two-point correlation properties of
 stochastic ``cloud processes''}
\author{Andrea Gabrielli$^{1,2}$}
\author{Michael Joyce$^3$} 
\affiliation{$^1$Statistical Mechanics and
Complexity Center, INFM-CNR, Physics Department, University ``La
Sapienza'' of Rome, P.le Aldo Moro 2, 00185-Rome, Italy\\
$^2$Istituto dei Sistemi Complessi, CNR, via dei Taurini
19, 00185-Rome, Italy\\
$^3$ Laboratoire de Physique
Nucl\'eaire et de Hautes Energies, UMR-7585,\\ 
Universit\'e Pierre et Marie Curie --- Paris 6,
75252 Paris Cedex 05, France}
\begin{abstract}
We study how the two-point density correlation properties of a 
point particle distribution are modified when each particle 
is divided, by a stochastic process, into an equal number of identical
``daughter'' particles. We consider generically that there may be
non-trivial correlations in the displacement fields describing 
the positions of the different daughters of the same ``mother''
particle, and then treat separately the cases in which there are, or 
are not, correlations also between the displacements of daughters 
belonging to different mothers. For both cases exact formulae are 
derived relating the structure factor (power spectrum) of the daughter 
distribution to that of the mother. These results can be considered as 
a generalization 
of the analogous equations obtained in \cite{displa} for the case 
of stochastic displacement fields applied to particle distributions. 
An application of the present results is that they give explicit
algorithms for generating, starting from regular lattice arrays,  
stochastic particle distributions with an arbitrarily high degree of 
large-scale uniformity. 
\end{abstract}
\pacs{02.50.-r, 61.43.-j, 98.80.-k}
\maketitle 

\section{Introduction}
\label{sec1}
Point processes, i.e., stochastic spatial distributions of
identical point particles, provide a very useful mathematical
scheme for many different $N-$body and granular physical 
systems such as crystals (both regular, and perturbed)
\cite{ziman,ashcroft,groma}, quasi-crystals \cite{radin}, structural
glasses, fluids \cite{hansen}, self-gravitating systems in astrophysics
and cosmology (see, e.g., \cite{millionbody, pee80, libro}). They find also 
many applications 
in a wide range of scientific fields:
computer image  problems \cite{cpu},
and bio-metrical studies \cite{renshaw} are only some examples of
systems which are usually represented as specific point processes with
appropriate spatial correlation properties.
The extension of knowledge about this class of stochastic
processes can therefore be of fundamental importance for the description 
and analysis of many scientific topics.  Indeed this is the reason why 
considerable mathematical effort has already been invested in the study
of this class of systems, and many useful results have been 
derived (e.g., see \cite{daley,kertscher,torquato98}).

In this paper we present the equations for the structure factor (or
power spectrum) of a point process obtained as follows. We start with
an arbitrary uniform spatial point particle distribution with a known
structure factor (SF).  We now suppose that each of these ``mother''
particle splits into a cloud of $m$ identical ``daughter'' particles,
where $m$ is a cloud-independent constant.  Each daughter particle is
then assumed to be displaced from its mother position by a stochastic
displacement which may, or may not, be correlated with the displacements
of other particles.  In other words each set of $m$ particles
initially lying at the same spatial point ``explodes'' forming a
``cloud'' of particles around it.  For this reason we call the point
process so generated a {\em cloud process}.

We suppose that the displacements applied to different particles
belonging to the {\it same} mother are symmetrically distributed with
arbitrary pair correlations.  One can choose, for instance, these
correlations in order to fix certain moments of the mass dispersion of
each cloud.  We will distinguish in the following between the two
cases in which the displacements applied to particles originating from
{\it different} mothers are, or are not, correlated. We note that the
results obtained for these cases are generalizations of those obtained
in \cite{displa}, where the case of a single daughter for each mother
was analyzed and solved exactly.

The main immediate application we have in mind of the present study,
and the one we discuss at some length, is the determination of the
constraints on the stochastic displacement field and the number of
daughters $m$ required to produce a target behavior of the SF 
at large scale (i.e. small wave number $k$) in the daughter
particle distribution.  More specifically we are interested in the
case of a large scale SF inherent to {\em
superhomogeneous} (or {\em hyper-uniform}) stochastic point particle
distributions \cite{glass,lebo,torquato}. This class of distribution is
characterized by the convergence of the SF to zero as $k
\rightarrow 0$. We will show explicitly that, for the case that the
mother distribution is a regular lattice array, the associated
(positive) exponent in the $k \rightarrow 0$ limit of the SF 
is related to the conservation of the local mass moments in
the passage from the single point particle to the cloud of daughter
particles\footnote{We note that approximate heuristic derivations of
some of these results can be found in the cosmological literature
(see, e.g., \cite{pee80, wandelt}) in which constructions of this kind
are considered in discussions of ``causality bounds'', i.e., bounds
imposed on the large scale behavior of the SF by the
existence of a causal horizon in cosmological models.}.

The paper is structured as follows. In the next section we introduce
our notation and the essential definitions, and then derive a general
expression for the SF of the daughter distribution, before averaging
over the realizations of the cloud process. In Sect.~\ref{sec3} we
derive our general result for the SF for the case that there are only
correlations between displacements of particles derived from the same
mother. In the following section we apply this result to the specific
case that the mother distribution is a regular lattice, and derive the
small $k$ behavior of the SF of the daughter distribution. In
Sect.~\ref{sec5} we then derive our result for the more general case
that there is also arbitrary correlation between the displacements of
daughters of different mothers, and then also consider the specific
case of an initial regular lattice.  As an example we derive the small
$k$ behavior of the SF of a lattice of correlated dipoles. In the
final section we summarize our results and discuss briefly both
possible developments and further applications of the results reported
here.

\section{Structure factor (SF) of a stochastic cloud process}
\label{sec2}

We start with a spatial distribution of $M$ ``mother'' particles in 
a cubic volume $V$, for which we write the microscopic particle density as: 
\be
n(\bx)=\sum_{i=1}^{M}\delta(\bx-\bx_i)\,.
\label{eq1}
\ee 
Let us denote by $\left<...\right>$ the average over the ensemble
of realizations of this point process, which we assume to be uniform
at large scales, i.e., to have a well defined (positive)
mean particle density $n_0$ in the limit $V\rightarrow \infty$.  The
SF (or power spectrum) is then defined (see, e.g., \cite{libro}) as
\be S_n(\bk)=\lim_{V\rightarrow\infty}\frac{\left<|\tilde
n(\bk;V)|^2\right>}{M} -(2\pi)^d n_0\delta(\bk)\,,
\label{eq2}
\ee
where the limit $V \rightarrow \infty$ is taken at fixed $n_0$, 
and 
\[\tilde n(\bk;V)=\int_Vd^dx\,n(\bx)e^{-i\bk\cdot\bx}=\sum_{i=1}^M
e^{-i\bk\cdot\bx_i}\]
is the Fourier transform (FT) of the point distribution in the volume $V$. 
Note that, defined in this way, $S_n(\bk \rightarrow \infty)=1$, 
which is the usual normalization of the shot noise present in any 
particle distribution at short distances\footnote{An alternative
quite widely used normalisation (e.g. in the cosmology literature)
differs by a factor of $1/n_0$.}. We suppose further that the point
distribution is statistically spatially homogeneous. In this case 
the SF is related by a FT to the usual two point correlation
function. More specifically, if we denote by $h_n(\bx)$ the 
off-diagonal part of the reduced two point correlation function, 
we have 
\begin{eqnarray}
&&\left<n(\bx_0)n(\bx_0+\bx)\right>-n_0^2=n_0\delta(\bx)+n_0^2h_n(\bx)\nonumber\\
&&=n_0\int \frac{d^dk}{(2\pi)^d}\,S_n(\bk)e^{i\bk\cdot\bx}\,.
\label{defs-cfn}
\end{eqnarray}
In other words
\[S_n(\bk)=1+n_0\tilde h_n(\bk)\]
with $\tilde h_n(\bk)=FT[h_n(\bx)]$.

Each particle in this distribution is a  ``mother'' point
which splits into $m\ge 1$  particles (``daughters'').
We take these latter to have identical unitary mass \footnote{We could
equally take each daughter to have a mass $1/m$, so that mass
is explicitly conserved. We choose unitary mass as this overall
normalization factor cancels out in the SF.}. 
The daughter particle distribution, which we call a {\em cloud
process}, thus clearly has an average particle density
$\rho_0=mn_0$.
\bef 
\includegraphics[height=6.5cm]{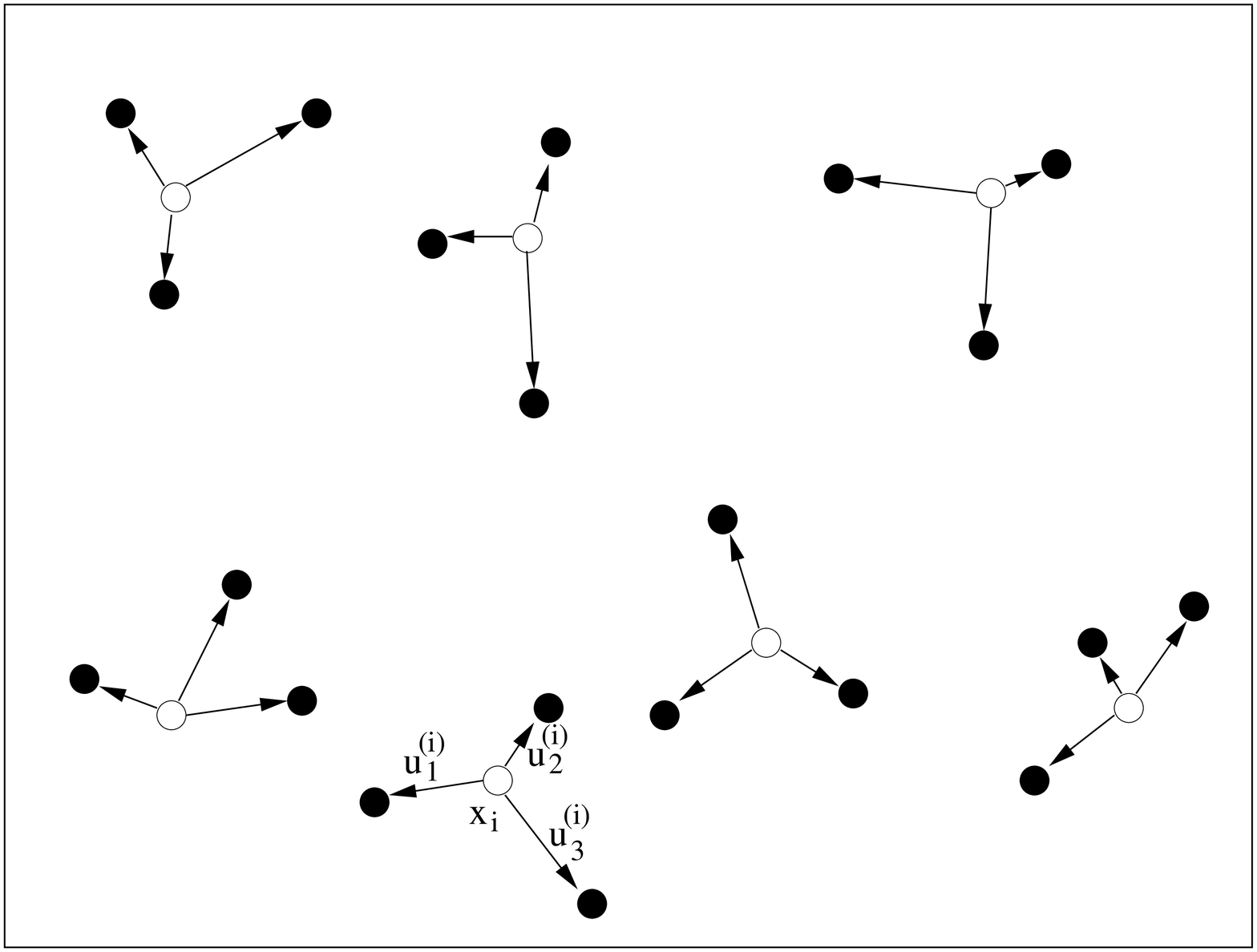}
\caption{The figure represents pictorially the generation of a
a ``cloud process'', as described in the
text and characterized by Eq.~(\ref{eq3}). Starting from a ``mother'' point 
process (white particles), a new particle distribution is generated by
the ``explosion'' of each mother particle into $m$ 
(here, $m=3$) daughters each of which is displaced from the original 
point by a stochastic displacement $\bu$. The problem we solve 
is the determination of the two-point correlation  properties of the 
new particle distribution given those of the ``mother'' distribution 
and the statistical properties of the displacement fields.
\label{fig1}}
\eef

Let us denote by $\buri$ the displacement from the mother position of
the $r^{th}$ particle ($1\le r\le m$) belonging to the cloud generated
by the $i^{th}$ mother ($1\le i\le M$) of the distribution $n(\bx)$.
The resulting microscopic density of particles is then (see
Fig.~\ref{fig1}) \be
\rho(\bx)=\sum_{r=1}^{m}\sum_{i=1}^{M}\delta(\bx-\bx_i-\buri)\,.
\label{eq3}
\ee
Let us now take the FT of $\rho(\bx)$ in the volume $V$
\[\tilde\rho(\bk;V)=\int_Vd^dx\,\rho(\bx)e^{-i\bk\cdot\bx}=
\sum_{r=1}^{m}\sum_{i=1}^{M}e^{i\bk\cdot(\bx_i+\buri)}\,.\]
We can now write the SF for $\rho(\bx)$ as
\be
S_{\rho}(\bk)=\lim_{V\rightarrow\infty}\frac{\left<\overline{
|\tilde\rho(\bk;V)|^2}\right>}{N}-(2\pi)^d\rho_0\delta(\bk)\,,
\label{eq4}
\ee where $\left<...\right>$ is, as above, the average over the
realizations of $n(\bx)$ and $\overline{(...)}$ is the average over
the realizations of the displacement field. We will always assume
here that the displacement field is statistically independent of 
the realization of the ``mother'' distribution, so that these 
two averages commute.  In general, we can write 
\bea
|\tilde\rho(\bk;V)|^2&=&N+\sum_{i=1}^{M}\sum_{r\ne
s}^{1,m}e^{-i\bk\cdot (\buri-\busi)}\nonumber\\ &+&\sum_{i\ne
j}^{1,M}\sum_{r, s}^{1,m} e^{-i\bk\cdot(\bx_i+\buri-\bx_j-\busj)}\,,
\label{eq5}
\eea
where $\sum_{r\ne s}$ denotes the sum over $r$ and $s$ excluding 
the diagonal terms  $r=s$, and similarly for $\sum_{i\ne j}$.

\section{Uncorrelated clouds: general result}
\label{sec3}

In this section we assume the following statistical properties for the 
displacement field:
\bei

\item The displacements applied to daughters with different mothers 
are statistically independent;

\item The displacements applied to different daughters with the same
mother may be arbitrarily correlated.

\eei

More precisely, let us denote by $p(\bu)$ the probability density
function (PDF) for a single displacement, by $p_s(\bu,\bv)$
the joint PDF of the displacements $\bu$  and $\bv$ applied to 
two daughters of the same mother, and by 
$p_d(\bu,\bv; \bx)$ the joint PDF of the
displacements applied to two particles belonging to different mothers
separated by $\bx$. Our assumptions imply that we have 
$p_d(\bu,\bv; \bx) = p(\bu) \, p(\bv)$ for any $\bx\ne {\bf 0}$, 
while we allow $p_s(\bu,\bv) \neq p(\bu) \, p(\bv)$. 

Note that in writing the PDFs in this way, without labels for the clouds 
and for the particles in a single cloud to which the displacements apply, 
we assume implicitly the following symmetries: 
(i) $p(\bu)$ is the same for all displacements, and (ii)
$p_s(\bu,\bv)$ does not depend on the cloud and is the same for all
$m(m-1)/2$ couples of the $m$ particles belonging to the same cloud.
In other words, if call $P(\bu_1,...,\bu_m)$ is the joint PDF of the
displacements applied to the $m$ daughter particles in a cloud, we
assume that $P(\bu_1,...,\bu_m)$ is the same for all clouds and 
is invariant under any permutation of the $m$ displacement variables.  
This implies in particular that $p_s(\bu,\bv)=p_s(\bv,\bu)$ and, 
as we show below, important constraints on the displacement-displacement 
correlations inside a single cloud.  
We clearly also have the consistency condition
\[ \label{consistency}
\int d^dv\,p_s(\bu,\bv)=p(\bu)\,.\]

With these assumptions it is simple to show that it follows from
Eq.~(\ref{eq5}) that 
\bea
\overline{|\tilde \rho(\bk)|^2}&=&N+N(m-1)\tilde p_s(\bk,-\bk)\nonumber\\
&+&m^2|\tilde p(\bk)|^2\sum_{i\ne j}^{1,M}e^{-i\bk\cdot(\bx_i-\bx_j)}\,,
\label{eq6}
\eea
where $\tilde p(\bk)=FT[p(\bu)]$ (i.e. the characteristic function of
the single stochastic displacement) and $\tilde p_s(\bk,-\bk)$
is the following diagonal two-point FT of $p_s(\bu,\bv)$:
\[\tilde p_s(\bk,-\bk)=\int d^du\, d^dv\,p_s(\bu,\bv)
e^{-i\bk\cdot (\bu-\bv)}\,. \]
Note that 
\[
\tilde p_s(\bk,-\bk)=\tilde \phi_s(\bk)=\int d^dw\,\phi_s(\bw)
e^{-i\bk\cdot\bw}\,,\] 
where $\phi_s(\bw)$ is the PDF of n-th relative displacement 
$\bw=(\bu-\bv)$ between two particles in the same cloud:
\[\phi_s(\bw)=\int d^du\,d^dv\,p_s(\bu,\bv)\delta(\bw-\bu+\bv)\,.\]
Since $p_s(\bu,\bv)=p_s(\bv,\bu)$, $\phi_s(\bw)$ is 
an even function of $\bw$,  and, as a consequence, $\tilde \phi_s(\bk)$ is 
a real function. Further, from the fact that $\phi_s(\bw)$ is a 
PDF, it follows that $\tilde \phi_s(\bk={\bf 0}) =1$, and 
$|\tilde \phi_s(\bk)|\le 1$ at all $\bk$.

To perform the average $\left<...\right>$ (over the
realizations of the mother distribution), we use that
\[\sum_{i\ne j}^{1,M}e^{-i\bk\cdot(\bx_i-\bx_j)}=
\sum_{i, j}^{1,M}e^{-i\bk\cdot(\bx_i-\bx_j)} -M\,,\] 
and that from Eq.~(\ref{eq2}), for asymptotically large $M$ (i.e. for
$V\rightarrow\infty$ with $n_0$ fixed), one has
\[\left<\sum_{i, j}^{1,M}e^{-i\bk\cdot(\bx_i-\bx_j)}\right>=
M[S_n(\bk)+(2\pi)^d n_0\delta(\bk)]\,.\]
Using these results, together with $N=mM$ and $\rho_0=mn_0$, 
we finally obtain
\be
S_\rho(\bk)=1+(m-1)\tilde\phi_s(\bk)-m|\tilde p(\bk)|^2+
m|\tilde p(\bk)|^2S_n(\bk)\,.
\label{eq7}
\ee
This is the principle result of this section. It is useful to write
it also in the form
\be
S_\rho(\bk)=S_\rho^{(0)}(\bk) + m|\tilde p(\bk)|^2S_n(\bk)\,.
\label{eq7-2pieces}
\ee
where 
\be
S_\rho^{(0)}(\bk)=1 + (m-1) \tilde \phi_s (\bk) -
m |\tilde p(\bk)|^2 
\label{first-term}
\ee
depends only on the statistical properties of the displacement fields 
(i.e. independent of those of the mother distribution). 

We note that for the case $m=1$ Eq.~(\ref{eq7}) reduces to
\be
S_1(\bk)=1-|\tilde p(\bk)|^2+|\tilde p(\bk)|^2S_0(\bk)\,.
\label{eq8}
\ee
This is precisely the equation derived in \cite{displa, libro} for the 
transformation of the PS of a point particle distribution from $S_0(\bk)$ 
to $S_1(\bk)$ when each  particle is randomly displaced, independently
of the others, with a PDF $p(\bu)$ for the single random displacements. 

Another simple case is that in which the mother particle distribution 
$n(\bx)$ is itself completely uncorrelated with $h_n(\bx)=0$, i.e., generated 
by a homogeneous Poisson process. Then $S_n(\bk)=1$ and 
consequently Eq.~(\ref{eq7}) gives
\[ S_\rho(\bk)=1+(m-1)\tilde \phi_s(\bk)\,,\]
Since $\tilde \phi_s(\bk \rightarrow 0)=1$ it follows that 
$S_\rho(\bk \rightarrow {\bf 0})=m$, i.e., at large scales the 
cloud process is identical to the original distribution (up
to a change in the mean particle density by the 
factor $m$). This simply translates the fact that if a
point process has no correlation, we cannot create correlation
at large scales by dividing the particles into clouds by a 
stochastic process which incorporates no correlation between
the different clouds.

\subsection{Properties of $S_\rho(\bk)$}

By definition any SF, and therefore the one we have derived, must satisfy 
the conditions
\bea
\label{prop1}
&&S_\rho(\bk) \geq 0 \\
&&S_\rho(\bk \rightarrow \infty)= 1 \,. 
\label{prop2}
\eea
These must hold for {\it any} input $S_n(\bk)$ (itself obeying these
properties) and any finite value of $m$. 

The second property Eq.~(\ref{prop2}) is simple to verify.  It follows
from the fact that both $\tilde\phi_s(\bk)$ and $|\tilde p(\bk)|$ 
vanish in the large $k$ limit. This is the case because they are FTs
of functions which are integrable at the origin.

That the first property Eq.~(\ref{prop1}) must be satisfied by our
result is trivial: we obtained $S_\rho(\bk)$ by simply averaging 
Eq.~(\ref{eq5}) which is non-negative definite by construction.  
However, as we now discuss, it is not simple as one might 
anticipate to verify it directly from Eq.~(\ref{eq7}).
The property Eq.~(\ref{prop1}) in fact encodes in a concise
and very non-trivial manner constraints on the 
joint PDF $p_s(\bu,\bv)$ which follow from the assumption that it is
unique for any two particles in a cloud.

Firstly we note that Eq.~(\ref{prop1}) holds in fact if and only if
\be
\label{condition-phi}
S_\rho^{(0)}(\bk) \geq 0
\ee 
for all $\bk$. This is the case because Eq.~(\ref{prop1}) must
be true for an arbitrary $S_n (\bk)$, and it is always possible to 
choose a mother point process for which it vanishes at any 
given $\bk$ (taking, e.g., an appropriate  regular lattice).
For $m=1$ this condition is trivially satified, as 
$|\tilde p(\bk)|\leq 1$ by definition (as FT of a PDF).
For $m\geq 2$ it may be rewritten as the condition
\be
 \tilde \phi_s (\bk) \geq -\frac{1}{m-1} + \frac{m}{m-1}|\tilde p(\bk)|^2 > -1
\label{constraint}
\ee
This does not trivially follow from the fact that $\tilde p(\bk)$ 
and $\tilde\phi_s(\bk)$ are FT of PDFs.  As noted above, the latter
gives only the weaker condition $\phi_s (\bk) \geq -1$.
Clearly Eq.~(\ref{constraint}) encodes a non-trivial constraint
on $\phi_s (\bk)$, arising from the fact that it is related to
a PDF for the {\it joint} displacement PDF. The latter is not 
simply an arbitrary normalizable function. The condition
Eq.~(\ref{constraint}) tells us that we have in fact
constrained it mathematically by the assumption about 
it we have made in our derivation: we have assumed that
the cloud is generated in such a way that the joint
two-displacement PDF is identical for all couples of particles.

To illustrate this more explicitly we derive now the form
taken by the constraint, in the form of Eq.~(\ref{condition-phi}),
when $S_\rho^{(0)}(\bk)$ is expanded in Taylor series around 
$\bk={\bf 0}$. Such an expansion can be made assuming that 
both $p(\bu)$ and $\phi_s(\bw)$ are rapidly decreasing at 
large $u$ and $w$ so that their FT are analytic at $\bk={\bf 0}$. 
We then have 
\be 
\tilde
p(\bk)=\sum_{l=0}^{\infty}(-i)^l
\frac{\overline{(\bk\cdot\bu)^l}}{l!}\,.
\label{eq11}
\ee
and 
\be
\tilde\phi_s(\bk)\equiv
\tilde p_s(\bk,-\bk)=\sum_{l=0}^{\infty}(-i)^l
\frac{\overline{(\bk\cdot\bw)^{l}}}{(l)!}\,.
\label{eq10}
\ee 

The condition (\ref{condition-phi}) can therefore be rewritten as
\bea
\label{cond-cpor}
S_\rho^{(0)}(\bk)=&&\sum_{l=1}^{\infty}\frac{(-i)^l}{l!}\left\{(m-1)
\overline{[\bk\cdot(\bu-\bv)]^l}\right.\\
&&\left.-m\sum_{l'=0}^{l}
{l\choose l'}\overline{(\bk\cdot\bu)^{l-l'}}\times
\overline{(\bk\cdot\bv)^{l'}}\right\}\ge 0\,,\nonumber
\eea
where we have written explicitly $\bw=(\bu-\bv)$ and
used the symmetry assumption 
$\overline{(\bk\cdot\bv)^{j}}=\overline{(\bk\cdot\bu)^{j}}$ for any $j$.
This inequality, valid for all $\bk$, fixes all the constraints
on the two-displacement correlation function in any cloud.

Using again the fact that $\phi_s(\bw)=\phi_s(-\bw)$, and 
making the further assumption that $p(\bu)=p(-\bu)$,
all the odd power terms in this expression vanish
so that we obtain:
\bea
\label{cond-cpor}
S_\rho^{(0)}(\bk)&=&\sum_{l=1}^{\infty}\frac{(-1)^l}{(2l)!}\left\{(m-1)
\overline{[\bk\cdot(\bu-\bv)]^{2l}}\right.\\
&&\left.-m\sum_{l'=0}^{l}
{2l\choose 2l'}\overline{(\bk\cdot\bu)^{2(l-l')}}\times
\overline{(\bk\cdot\bv)^{2l'}}\right\}\ge 0\,.\nonumber
\eea
The leading term dominates at sufficiently small $k$ and 
therefore has to be non-negative. This implies the 
following constraint on the correlations of any 
two displacements in the set of $m$ correlated 
random displacements in each cloud: the matrix
\be
c_{\mu\nu}\equiv (m-1)\overline{u^{(\mu)} v^{(\nu)}}
+\overline{u^{(\mu)} u^{(\nu)}}\,,
\label{matrix}
\ee
where $\mu,\nu=1,...,d$, has to be non-negative definite.
In our analysis in the next section, of the case that 
the mother distribution is a lattice, we will see how this
constraint, and ones which can derived at subsequent
order in this expansion, simplify and are explicitly
verified in certain cases.

\subsection{Behaviour at small $k$}

Let us now consider specifically the properties of the SF
of the mother and daughter distributions as $\bk \rightarrow 0$.

We note first that, because of the normalization conditions on 
the PDFs of the displacements, both $\tilde\phi_s(\bk)$ and 
$|\tilde p(\bk)|$ converge continuously to unity as 
$\bk \rightarrow {\bf 0}$. It follows from Eq.~(\ref{first-term})
that $S_\rho^{(0)}(\bk \rightarrow {\bf 0}) \sim k^{\alpha}$, where 
$\alpha > 0$. Supposing now that the initial (mother) point 
distribution has $S_n(k \rightarrow 0) \sim k^\gamma$, we
can infer that $S_\rho (k \rightarrow 0) \sim k^{\gamma'}$ where
(i) $\gamma' = \gamma$ for $\gamma \leq 0$, and (ii)
$\gamma'= min \{\gamma, \alpha \}$ for $\gamma > 0$. 

Thus the exponent of the SF around $k=0$ {\it can never be larger} in
the cloud process than in the original mother point process. Further
it may differ from it (in which case it is smaller) only if
$S_n(k=0)=0$. Note that these conclusions hold independently of any
assumption about the cloud process, other than that there is no
correlation between the displacement sets creating different clouds
and that the displacements are symmetrically distributed as shown
above.

This result may be explained more physically as follows. The exponent
of the small $k$ behavior of the SF can be considered as a measure of
the degree of order in the stochastic point process at asymptotically
large scales \cite{glass, torquato}. The greater the exponent the more 
ordered is the distribution. Indeed any lattice, which is the class 
of the most ordered particle distributions, the SF vanishes identically around
$\bk={\bf 0}$, i.e., we can consider it to correspond to the behaviour
$~k^\infty$. Clearly a cloud process, without any correlation between
the arrangement of matter in the different clouds, cannot increase the
degree of order.
That it may,
on the other hand, decrease the degree of order when the mother
distribution has the property that $S_n(k=0)=0$, reflects the
difference between this class of distributions and those with
$S_n(k=0) >0$. Indeed this difference is that underlined by the
classification of the former point processes as {\it
super-homogeneous} (or {\it hyper-uniform}): the rapid decay of the
density fluctuations at long wavelengths which characterizes them are
the result of a delicate balance between small scale and large scale
correlations in direct space. Indeed the condition $S_n(k=0)=0$ is
explicitly an integral constraint on the two point correlation
function over all space.  The processes which we are considering, in
which each particle ``explodes'' independently of all others, can
break these global constraints by modifying only the small scale
correlation properties. Instead for distributions with $S_n(k=0)\neq
0$\footnote{These can be classified into Poisson-like for $S_n(k=0) <
+\infty$, and long-range correlated for $S_n(k\to 0) \to +\infty$.}, which
do not present such a correlation balance,
such an uncorrelated re-distribution of matter at small scales cannot
modify the nature of the system at large scales and also.

Let us now consider the problem of the construction of a point process
with a target behavior of $S(\bk \rightarrow {\bf 0})$. From the results we
have just derived it follows that a cloud process of the type we have
just analyzed, without correlations between the displacements of the
members of different clouds, may be used to generate a distribution
with the target exponent $\alpha > 0$ provided we start with $\gamma >
\alpha$. In practice we can start with $\gamma=\infty$ by taking a
regular lattice, for which $S_n(\bk)=0$ identically in a finite region
around $k=0$ (specifically, in the first Brillouin zone). The
generated distribution will then have the exponent $\alpha$, which
depends through Eq.~(\ref{first-term}) only on the statistical
properties of the displacement fields. Since $\alpha>0$ the generated
process is necessarily superhomogeneous. The question we now address
is what values of $\alpha$ are attainable, and for what conditions on
the number of daughters $m$ and the displacement fields they are
realized. We note that in \cite{displa} it has been shown that, for
$m=1$, one can realize by appropriate choice of $p(\bu)$ any exponent
in the range $0<\alpha \leq 2$. The upper bound $\alpha=2$ results for
any $p(\bu)$ with a finite variance, while the lower exponents are
realized for PDFs with appropriately divergent moments of
displacements at order less than two. For the case of correlated
displacements, with Gaussian statistics, it has been shown also in
\cite{displa} (see also \cite{bruno-mike}) that a maximal value of
$\alpha=4$ may be attained. We will see now that, with an appropriate
value of $m$ and conditions on the displacement fields, arbitrarily
large positive target value of $\alpha$ are attainable. We will show
that to obtain a certain value of $\alpha$ requires that one fixes a
sufficiently large number of mass moments of all clouds of particles
with respect to their respective initial position on a regular
lattice.


\subsection{Explicit expansion around $k=0$}

For this study of the small $k$ behavior of 
$S_\rho(\bk)$ we consider the expansion of 
both $\tilde \phi_s(\bk)$ and $\tilde p(\bk)$
in power series of $\bk$, which are given respectively 
by Eqs.~(\ref{eq10}) and (\ref{eq11}).
Using these expressions in Eq.~(\ref{eq7}) we obtain 
\bea
S_\rho(\bk)&=&1+(m-1)\sum_{l=0}^{\infty}(-i)^l
\frac{\overline{[\bk\cdot(\bu-\bv)]^l}}{l!}\nonumber\\
&+&m\left|\sum_{l=0}^{\infty}(-i)^l
\frac{\overline{(\bk\cdot\bu)^l}}{l!}\right|^2[S_n(\bk)-1]\,.
\label{eq12}
\eea 
All these expressions are valid only under the assumption that
all the moments in the sums are finite, while the derivation of
Eqs.~(\ref{eq7}) and (\ref{first-term}) required only the
integrability of the probability distributions.  If the probability
distributions $p(\bu)$ and $\phi_s(\bw)$ have only a finite number of
finite moments, the corresponding sums must be terminated at the
appropriate order. There is then an additional term, of which the
leading singular part can easily be determined. We will not discuss
here the case in which there are such singularities \footnote{We will
see below that to obtain exponents greater than four with uncorrelated
clouds the PDFs for the displacements must, in fact, have compact
support.}.  The required generalization of the analysis described here
is straightforward following the procedure defined in \cite{displa}.

\section{Uncorrelated cloud lattice}
\label{sec4}

Given the above motivation we now analyze in detail the cloud
processes in the previous section for the particular case 
that the mother particle distribution is a regular
lattice, i.e.
\[n(\bx)=\sum_{\bR}\delta(\bx-\bR)\,,\]
where $\bR$ is the generic lattice site.
In this case the SF of $n(\bx)$ is
\be
S_n(\bk)=(2\pi)^dn_0\sum_{\bH\ne 0}\delta(\bk-\bH)\,,
\label{lattice-ps}
\ee
where the sum is over all the vectors $\bH$ of the reciprocal lattice
but $\bH={\bf 0}$.  Note that this vanishes identically in the first
Brillouin zone, and therefore in this region of the $\bk-$space the
following relation holds exactly:
\be
S_\rho(\bk)=1+(m-1)\sum_{l=0}^{\infty}(-i)^l
\frac{\overline{(\bk\cdot\bw)^l}}{l!}-m\left|\sum_{l=0}^{\infty}(-i)^l
\frac{\overline{(\bk\cdot\bu)^l}}{l!}\right|^2\,.
\label{eq13}
\ee 
To simplify the presentation of our determination of the conditions 
required to have an arbitrary (analytic) small $k$ behavior of
$S_\rho(\bk)$,  we start
with the one-dimensional case, for which Eq.~(\ref{eq13}) becomes 
\be
S_\rho(k)=1+(m-1)\sum_{l=0}^{\infty}(-ik)^l
\frac{\overline{(u-v)^l}}{l!}-m\left|\sum_{l=0}^{\infty}(-ik)^l
\frac{\overline{u^l}}{l!}\right|^2\,,
\label{eq14}
\ee
where $u$ and $v$ are the displacements applied to two different 
particles belonging to the same cloud.
First of all we see immediately, as above underlined,
that the final particle distribution $\rho(x)$ is superhomogeneous, as 
the zero order contribution in $k$ to $S_\rho(k)$ vanishes identically 
for any choice of $p_s(u,v)$ [i.e. of $\phi_s(w)$]. In
the notation of the previous section, we have explicitly that
$\alpha \geq 1$.  We can write
\[\left|\sum_{l=0}^{\infty}(-ik)^l\frac{\overline{u^l}}{l!}\right|^2
=\sum_{l=0}^\infty
\frac{(-ik)^l}{l!}  \sum_{j=0}^{l}(-1)^{j}{l\choose j}
\overline{u^{j}}\times\overline{v^{l-j}}\,,\]
where we have used $u$ and $v$ instead of only $u$ as the moments of the 
single displacement are the same for every particle.
Moreover, by expanding the terms $(u-v)^l$ in Eq.~(\ref{eq14}), we have
\[\sum_{l=0}^{\infty}(-ik)^l\frac{\overline{(u-v)^l}}{l!}=
\sum_{l=0}^\infty
\frac{(-ik)^l}{l!}  \sum_{j=0}^{l}(-1)^{j}{l\choose j}
\overline{u^{j}\times v^{l-j}}\,.\]
Therefore Eq.~(\ref{eq14}) becomes
\bea
&&S_\rho(k)=\sum_{l=1}^{\infty}\frac{(-ik)^l}{l!}
\sum_{j=0}^{l}(-1)^{j}{l\choose j}
\times\nonumber\\
&&\left[(m-1)\overline{u^{j}\times v^{l-j}}-
m\,\overline{u^{j}}\times\overline{v^{l-j}}\right]
\label{eq15}
\eea
Making the additional assumption of statistical symmetry 
in the displacements, $p(u)=p (-u)$, 
all the terms with odd $l$ in Eq.~(\ref{eq15}) vanish.

\subsection{Order by order analysis and conservation of mass moments ($d=1$)}
\label{4a}

Let us now analyze in detail Eq.~(\ref{eq15}), denoting by ${\cal O}_n(k)$ 
its term proportional to $k^n$. Given our hypotheses the lowest order
non-zero term is $n=2$:
\[{\cal O}_2(k)=\left[\overline{u^{2}}+(m-1)\overline{u\times v}\right]
k^2\,.\] 
It is simple to verify explicitly that
$\left[\overline{u^{2}}+(m-1)\overline{u\times v}\right]\ge 0$ always,
as required (from the fact that $S_\rho(k)$ is a SF).
First of all it is the one-dimensional version of the condition
(\ref{matrix}).  This can be seen more directly as follows: if we denote by
$u_i$ with $i=1,...,m$ the displacements applied respectively to the
$m$ daughter particles originating from the same mother
(i.e. belonging to the same cloud), it is clear that
\[\overline{\left(\sum_{i=1}^m u_i\right)^2}\ge 0\,.\]
This quantity, however, given our symmetry hypotheses about the displacments
distribution is nothing other than 
\[ \overline{\left(\sum_{i=1}^m u_i\right)^2}=
m\left[\overline{u^{2}}+(m-1)\overline{u\times v}\right]\,.\] 

Consequently the condition to have an identically vanishing ${\cal O}_2(k)$
term, and therefore a small $k$ SF of order greater
than two (i.e. $\alpha > 2$), is 
$\overline{\left(\sum_{i=1}^m u_i\right)^2}=0$, or in other words, 
\be
\sum_{i=1}^m u_i=0
\label{unmoved-com}
\ee 
with probability one\footnote{Note that if we had allowed an
asymmetric $p(u)$ with a non-zero average value $\overline{u}$ the
same condition ${\cal O}_2(k)=0$ would have been written as
$\sum_{i=1}^m u_i=m\overline{u}$ for all the clouds.}. This means that
the center of mass of each cloud does not move away from the mother
particle when the displacements are applied.  Clearly for $m=1$ this
condition can only be trivially satisfied by applying no displacement,
in which case the daughter distribution is the original lattice
distribution.  For $m=2$ it can be satisfied non-trivially: choosing
the displacement of a first point with the PDF $p(u)$, the other
particle is then displaced deterministically by $-u$.  For $m>2$ the
condition can be satisfied while admitting a higher degree of
stochasticity: it fixes deterministically the displacement of only one
particle among $m$ once the other $(m-1)$ are chosen stochastically.

The analysis of the term ${\cal O}_4(k)$ is more complex. 
We will now show that the condition for it to vanish is one on 
the second moment of the mass dispersion of each cloud.  
Directly from Eq.~(\ref{eq15}) we have 
\be 
{\cal O}_4(k)=-\frac{k^4}{12}
\left[\overline{u^4} 
+4(m-1)\overline{u^3v}-3(m-1)\overline{u^2v^2}+3m\sigma^4 \right]\,,
\label{eq17}
\ee 
where we have denoted $\sigma^2=\overline{u^2}$ and again used the
assumed symmetries of the displacement field. If the term of ${\cal
O}_2(k)$ vanishes at all $\bk$, i.e., if Eq.~(\ref{unmoved-com}) is
satisfied for each cloud, then the ${\cal O}_4(k)$ term must be
non-negative [since the SF $S_\rho(k)$ must be non-negative at all
$k$]. Thus the coefficient of $k^4$ in Eq.~(\ref{eq17}) must be
non-negative. In order to show that this is indeed the case we note
first that Eq.~(\ref{unmoved-com}) implies that
\[\overline{u_1^3\sum_{i=1}^m u_i}=\overline{u^4}+(m-1)\overline{u^3v}=0\,.\]
By using this relation in the coefficient of Eq.~(\ref{eq17}) we 
then have
\[
\begin{array}{l}
-\overline{u^4}-4(m-1)\overline{u^3v}+3(m-1)\overline{u^2v^2}-3m\sigma^4\\
=3\overline{u^4}+3(m-1)\overline{u^2v^2}-3m\sigma^4\,.
\end{array}
\]
Further it is simple to show that 
\bea
\label{eq18}
&&3m[\overline{u^4}+(m-1)\overline{u^2v^2}-m\sigma^4]\\
&&=3\left[\overline{\left(\sum_{i=1}^mu_i^2\right)^2}
-\left(\overline{\sum_{i=1}^m
u_i^2}\right)^2\right]\,,\nonumber
\eea 
which is manifestly non-negative.
 
This result also gives the condition necessary in order
to have both the ${\cal O}_2(k)$ and ${\cal O}_4(k)$ identically
vanishing: once Eq.~(\ref{unmoved-com}) for the conservation of the
center of mass of each cloud of the system is assumed, in order 
to make the variance  in Eq.~(\ref{eq18}) vanish one requires 
also that
\be
\sum_{i=1}^mu_i^2=m\sigma^2
\label{eq19}
\ee 
for each cloud, with probability one.  

In summary: in order to obtain with this algorithm a particle
distribution with a SF of order larger than 
$k^4$ at small $k$, one has to satisfy exactly
the following two conditions:
\begin{itemize}
\item every cloud must have the same displacement of its
center of mass from the initial point of the cloud or mother position
[and, in particular, if $p(u)=p(-u)$ the center of mass of the cloud
must coincide with the position of the mother particle];
\item every cloud must have the same inertial moment (or second
moment of its mass dispersion) with respect to the
initial point. The value of this inertial moment is fixed by the
second moment of $p(u)$ and $m$ as $m\sigma^2$.
\end{itemize}
This analysis can be continued in order to determine the 
conditions needed to obtain a $S_\rho(k)$ of order higher 
than $k^{2n}$ at small $k$, for any integer $n$. The result 
is simply that in order to obtain this goal one 
has to fix the first $n$ moments of the mass dispersion 
\be 
\sum_{i=1}^m
u_i^j=c_j\;\;\mbox{with}\;j=1,...,n
\label{eq20}
\ee 
where the constants $c_j$ are determined by the $j$-th moment of 
$p(u)$ and $m$ as $c_j=m\overline{u^j}$.  Clearly this gives $n$
conditions and consequently one has to have at least $m=n$ particles 
in each cloud in order to make the requirements given by Eq.~(\ref{eq20})
realizable. For $m=n$ there may be a single non-trivial solution to
the constraints, i.e., a unique choice of displacements. In this case
the generated distribution will be a lattice with basis, with
a SF which again vanishes in a finite region around
$k=0$. For $m>n$ the set of constraints may be satisfied (for
some range of values of the constants $c_j$) while leaving free
$(m-n)$ degrees of freedom. These may be then fixed stochastically,
leading generally to a stochastic particle distribution 
\footnote{Here we mean by ``stochastic" that there is a
non-empty compact domain of  $\bk-$space in which $S_\rho(\bk)$ is
continuous and strictly positive.} with a leading non-zero term 
at ${\cal O}_{2(n+1)}(k)$.
For $n=2$, for instance, we need at least two particles. In order to
fix the center of mass of the pair at the lattice site, and its 
second mass moment to a given value $c_2$, clearly fixes deterministically 
the points to lie at $\pm \sqrt{c_2/2}$. Taking three particles
one can instead satisfy the constraints fixing one degree of freedom 
stochastically: placing one point at $u$ with probability $p(u)$, the
position $u'$ of a second point is determined by solving the 
quadratic equation $(u')^2 + uu' + (u^2-c_2/2)=0$, and finally 
a third point is placed at $-(u+u')$. Note that the existence
of a solution to the quadratic equation places a strict upper bound
on $u$, $u \leq \sqrt{2c_2/3}$. Thus the probability $p(u)$ 
necessarily has finite support (and cannot in particular be
Gaussian) which is proportional to $\sqrt{c_2}$. 
It is clear that this is a general 
requirement for any algorithm of this kind producing a SF
with a leading small $k$ behavior $k^n$ with $n>4$:
in order to make the coefficient of the $k^4$ term vanish
the second moment of the mass dispersion of the cloud must 
be limited with probability one. 
Since the displacement of any particle contributes in
proportion to its square, the probability of displacement
larger than $\sqrt{c_2}$ must be zero. 

\subsection{Generalization to $d>1$}

In dimensions higher than one the problem is essentially 
the same. The analysis is, however, considerably more complicated 
because of the vectorial nature of the displacements. 

Let us first consider the conditions required to make the terms 
${\cal O}(k^{2})$ and ${\cal O}(k^{4})$ vanish. Fixing the center of 
mass in $d$ dimensions gives $d$ scalar
equations
\[\sum_{i=1}^m u_i^{(\mu)}=0\]
where $u_i^{(\mu)}$ is the $\mu^{th}$ (with $\mu=1,...,d$) component
of the displacement of the $i^{th}$ particle of the cloud.  To satisfy
this condition non-trivially evidently requires that there are at least
two particles in each cloud. Fixing the second moments of the mass
dispersion of the cloud gives $d(d+1)/2$ scalar equations, i.e., $d$
equations of the form
\[\sum_{i=1}^m \left[u_i^{(\mu)}\right]^2=a_{\mu\mu}>0\;\;\mbox{for}\;
\mu=1,...,d\]
with $a_{\mu\mu}=m\overline{[u_i^{(\mu)}]^2}$, and $d(d-1)/2$ equations of
the form
\[\sum_{i=1}^m u_i^{(\mu)}u_i^{(\nu)}=a_{\mu\nu}\;\;
\mbox{for}\;1\le\mu < \nu\le d\,,\] 
with $a_{\mu\nu}=m\overline{u^{(\mu)}u^{(\nu)}}$.
Therefore
to obtain a SF $S_{\rho}(\bk)$ of order larger than 
the fourth at small $k$ imposes $[d+d(d+1)/2]=d(d+3)/2$ scalar 
constraints on the displacements. This counting of constraints
may be continued to  higher orders, determining the number of 
conditions ${\cal N}(n,d)$ which must be satisfied to 
obtain, in $d$ dimensions, 
an $S_\rho(\bk)$ vanishing faster than $k^{2n}$ at small $k$.
Noting that all the moments of given order, say $l$, of the mass 
dispersion constitute a fully symmetric $l$-rank tensor in $d$ 
dimensions, which has ${d+l-1\choose l}$ independent components,
we find
\[{\cal N}(n,d)=\sum_{l=1}^{n}{d+l-1\choose l}\,.\]
Generalizing the reasoning for the case $d=1$, one might then 
be tempted to conclude that, since each particle 
brings $d$ degrees of freedom, the minimal number $m$ of particles
per cloud required to given a $S_\rho(\bk)$ of order
larger than $k^{2n}$ at small $k$ is ${\cal N}(n,d)/d$.
This conclusion is, however, not correct in that it tells us the
number of particles required to satisfy such conditions for {\it
arbitrary} physical values of the mass moments. To make the
coefficients in the expansion of $S_\rho(\bk)$ vanish up to some
order, while remaining non-vanishing at subsequent orders, we require
only that a number of particles sufficient to allow us fix {\it a set}
of physical values of the mass moments up to a certain order, while
allowing higher moments to vary. Put another way, by imposing
additional symmetries or constraints on the PDF of the displacements,
we can reduce the number of non-trivial constraints (i.e. equations),
reducing the others to simple identities. The following example
illustrates this point trivially: in $d>1$ we can always make a cloud
lattice by putting together at infinite number of one dimensional
cloud lattices, i.e., by constraining the displacements of the particles
to lie along a chosen axis of the lattice.  The calculation in $d=1$
then remains valid, as all the additional constraints on the mass
moments of the clouds with components in orthogonal directions are
trivially satisfied. The same is true in fact if the displacements are
in an arbitrary (but fixed) direction. The number $m$ of particles per
cloud required to obtain a SF with given leading order then remains
the same as in $d=1$. One can also evidently consider less radical
``dimensional reductions'', taking in $d$ dimension the displacements
of particles in each cloud only in a hyperplane of dimension smaller
than $d$.  \bef \includegraphics[height=6.5cm]{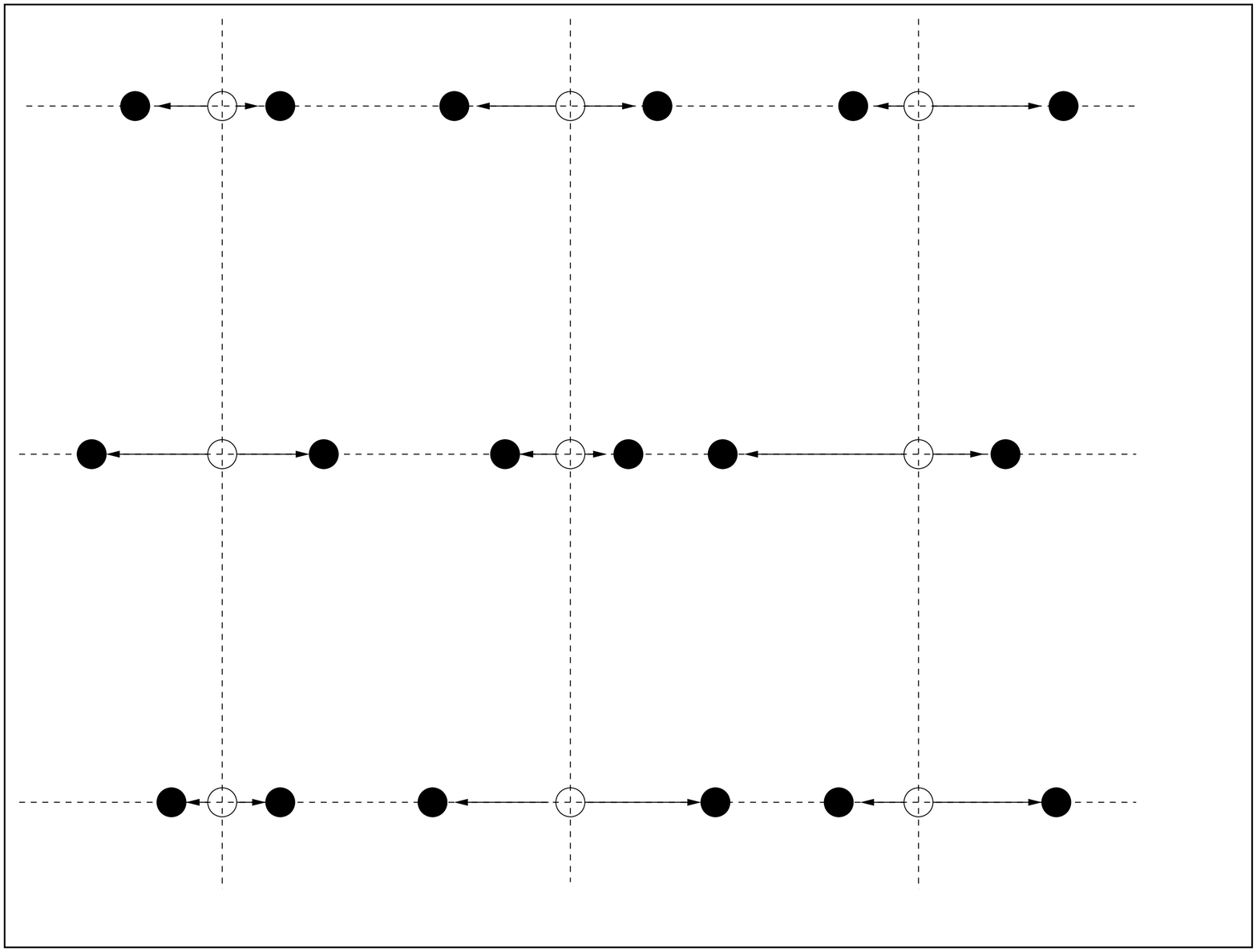}
\caption{The figure represents pictorially the generation of a
a cloud process starting from a lattice,  in $d=2$ with a 
``dimensional reduction'' to $d=1$ in which the 
two particles in each cloud are displaced along the same
direction.
\label{fig2}}
\eef

Even without such a reduction to a lower dimensional problem, it is
easy to give examples in $d>1$ which satisfy the constraints required
to make all terms of the SF up to a certain order $2n$ vanish, with
much less than ${\cal N}(n,d)/d$ particles, and which do not have the
feature of the above examples of breaking the statistical spatial
isotropy of the clouds. Consider, for example, the case that each
cloud contains an {\it even} number of particles, arranged
symmetrically with respect to the center of mass.  From the derivation
we have given above, it follows immediately that $S_\rho(\bk) \propto
k^\gamma$ at small $k$ with $\gamma \geq 4$ (as the inversion symmetry
gives automatically the conservation of the centre of mass). As shown
above the coefficients of the term proportional to $k^4$ will vanish
if the second order moments of the mass distribution are
cloud-independent.  This can be attained, for example, in $d$
dimensions, by taking $d$ pairs of particles arranged symmetrically
about, and equidistant, from the origin, and all mutually
orthogonal. This gives a second moment of the mass dispersion of the
cloud which is proportional to the identity matrix, and therefore
invariant under a random rotation $R \in SO(d)$ of the whole
configuration . Further it is possible to show (see Appendix A for
detail) that, because of the imposed inversion symmetry, the terms
proportional to $k^6$ vanish identically. Thus, placing such a cloud
with an orientation chosen randomly at each site, one obtains a
leading non-zero term of order $k^8$. This term is non-zero because
the (tensor) fourth mass moment is not invariant under rotation of the
configuration.  Further, if the stochastic process determining the
orientation is statistically isotropic, the SF at small
$k$ reflects this isotropy and is a function of $k$ only (rather than
the vector $\bk$).  It is simple to check that the number of required
particles for this algorithm $2d$ is less, for any $d$, than the
number ${\cal N}(3,d)/d$ given above.  \bef
\includegraphics[height=6.5cm]{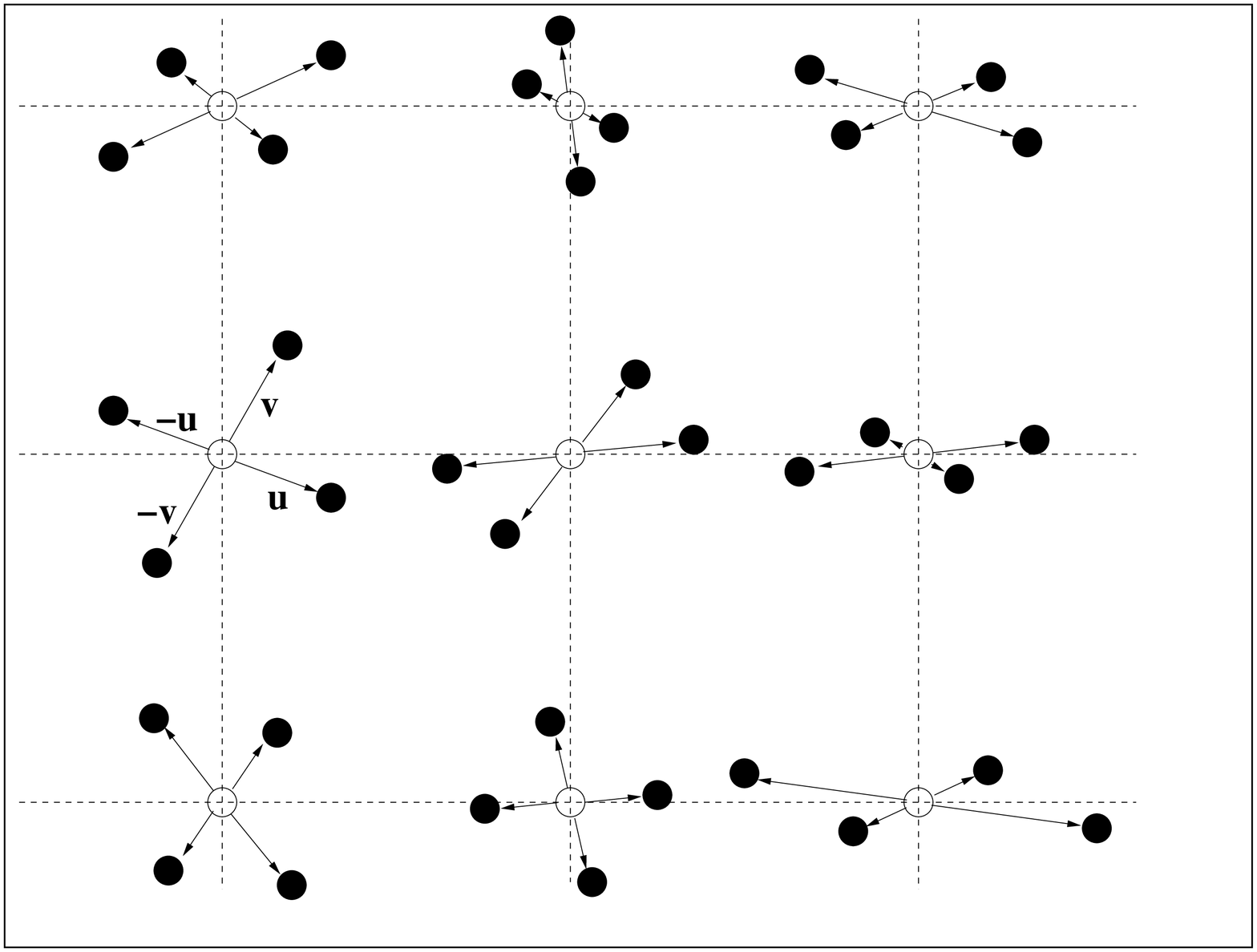}
\caption{The figure represents pictorially the generation of a
a lattice ``cloud process'', 
in $d=2$, in which each cloud is symmetric with respect to its
lattice site. 
\label{fig3}}
\eef


\section{Correlated clouds}
\label{sec5}

We now consider the case in which also displacements applied to
particles belonging to different clouds may be arbitrarily correlated.
In the notation introduced above for the joint PDFs of the 
displacements, this means that we now assume, at least for some
separation of mother particles $\bx$, that  
$p_d(\bu,\bv, \bx) \ne p(\bu)p(\bv)$.  
Further we make the following natural assumptions:
\bea
&&p_d(\bu,\bv;\bx)=p_d(\bv,\bu;-\bx) \nonumber\\ 
&&\lim_{x\rightarrow\infty} p_d(\bu,\bv;\bx)=p(\bu)p(\bv)\,.
\label{eq22}
\eea 
We have, of course, also the consistency condition
\be 
\int
d^dv\,p_d(\bu,\bv;\bx) = p(\bu)\;\;\forall \bx 
\ee
Note also that, strictly speaking, the displacement field is 
not continuous as a function of spatial separation because 
the correlation
$\overline{\bu\cdot\bv}(\bx)$ between two displacements applied to two
spatial point at vector distance $\bx$ in general does not converge to
$\overline{u^2}$ for $\bx\rightarrow {\bf 0}$. The natural choice for this
limit is $p_d(\bu,\bv;\bx\rightarrow
{\bf 0})=p_s(\bu,\bv)$. How precisely this limit is taken will not,
however, be of importance for our main application concerning
a regular lattice distribution of the centers of the clouds,
as in this case the distance between two different
centers is always different from zero.

In order to find the expression for $S_\rho(\bk)$, once
$S_n(\bk)$ and $p_d(\bu,\bv;\bx)$ are given, we have to go back to
Eq.~(\ref{eq5}). From this it is simple to show that
\bea
\frac{\overline{\left|\tilde\rho(\bk)\right|^2}}{N}&=&1+(m-1)
\tilde p_s(\bk,-\bk)\nonumber\\
&+&\frac{m^2}{N}\sum_{i\ne j}^{1,M}
e^{-i\bk\cdot(\bx_i-\bx_j)}\tilde p_d(\bk,-\bk;\bx_{ij})\,,\nonumber
\eea
where $\bx_{ij}=\bx_{i}-\bx_{j}$ and 
\[\tilde p_d(\bk,-\bk;\bx)=\int d^du\,d^dv\,p_d(\bu,\bv;\bx)
e^{-i\bk\cdot(\bu-\bv)}\,.\] 
Note that, analagously to $\tilde p_s(\bk,-\bk)$, 
the function $\tilde p_d(\bk,-\bk;\bx)$ is
the characteristic function of the stochastic vector
$\bw(\bx)=\bu(\bx)-\bu({\bf 0})$, i.e., of the difference between the
displacements applied to two particles belonging to clouds whose mother
particles are separated by $\bx$. 
In other words, if $\phi_d(\bw;\bx)$ is the PDF of $\bw(\bx)$, we have
\[\tilde p_d(\bk,-\bk;\bx)=\tilde\phi_d(\bk;\bx)\,,\]
where $\tilde\phi_d(\bk;\bx)=FT[\phi_d(\bw;\bx)]$.
In order to perform the average $\left<...\right>$, we recall that in
general for any function $f(\bx,\by)$ of two spatial variables one has
\[\left<\sum_{i\ne j}^{1,M} f(\bx_i,\bx_j)\right>=n_0^2\int\!\!\int_V
d^dx\,d^dy\,f(\bx,\by)\left[ 1+h_n(\bx-\by)\right]\] 
where $h_n(\bx)$ is the two point correlation function as defined
in Eq.~(\ref{defs-cfn}). Using this relation, we obtain
\bea &&\left<\sum_{i\ne
j}^{1,M}e^{-i\bk\cdot(\bx_i-\bx_j)}\tilde
\phi_d(\bk;\bx_{ij})\right>= n_0^2\int\!\!\int_V
d^dx\,d^dy\times \nonumber\\ &&e^{-i\bk\cdot(\bx-\by)}\tilde
\phi_d(\bk;\bx-\by) \left[ 1+h_n(\bx-\by)\right]\,.
\label{eq23}
\eea 
Taking the limit $V\rightarrow \infty$ with
$N/V=\rho_0$ and $m\ge 2$ fixed, we arrive at our result:
\bea
\label{eq24}
&&S_\rho(\bk)=1+(m-1)\tilde \phi_s(\bk)
\\
&&+\rho_0\int
d^dx\,e^{-i\bk\cdot\bx}\tilde \phi_d(\bk;\bx)\left[1+h_n(\bx)\right]
-(2\pi)^d\rho_0\delta(\bk)\,.\nonumber
\eea
Note that the case $m=1$ can be included in Eq.~(\ref{eq24}) by
considering, only for this value of $m$, $p_d(\bu,\bv;\bx)$
to be spatially continuous, i.e., converging to $p(\bu)\delta(\bu-\bv)$
for $\bx\rightarrow {\bf 0}$. The result  Eq.~(\ref{eq24}) for $m=1$
then agrees exactly with the analagous equation found 
in \cite{displa} for the PS of a particle distribution after 
the application of a correlated displacement field.

We will not analyse here the small $\bk$ expansion of Eq.~(\ref{eq24}),
but simply note that such an analysis can be done easily
by following the steps for the study of the similar
equation for the case $m=1$ in \cite{displa}.
 
Let us now give the special expression of Eq.~(\ref{eq24}) for the
specific case in which $n(\bx)$ is a regular lattice point
distribution.

\subsection{Correlated cloud lattice}
\label{sec5a}

When $n(\bx)$ is a regular lattice
\[h_n(\bx)={1\over n_0} \sum_{\bR\ne {\bf 0}} \delta(\bx-\bR)-1\,,\] 
where $\bR$ runs over
all the lattice vectors except $\bR={\bf 0}$. We therefore have
\bea
\label{eq25}
S_\rho(\bk)&=&1+(m-1)\tilde \phi_s(\bk)\\
&+&m \sum_{\bR\ne {\bf 0}}e^{-i\bk\cdot\bR}
\tilde \phi_d(\bk;\bR)-(2\pi)^d\rho_0\delta(\bk)\,.\nonumber
\eea 
This formula is a good starting point for the study of 
the small $\bk$ behavior of $S_\rho(\bk)$ for a cloud lattice 
for different choices of $p_d(\bu,\bv;\bR)$ and $m$.  We now use 
the following chain of
identities to rewrite Eq.~(\ref{eq24}) in a more useful form: 
\bea
&&(2\pi)^d\rho_0\delta(\bk)=(2\pi)^d\rho_0\left[\sum_{\bH}
\delta(\bk-\bH)-\sum_{\bH\ne {\bf 0}}
\delta(\bk-\bH)\right]\nonumber\\
&&=m \sum_{\bR}e^{-i\bk\cdot\bR}-mS_n(\bk)\,,
\label{eq26}
\eea
where we have used the definition (\ref{lattice-ps}) of $S_n(\bk)$ for
a regular lattice and the lattice identity
\be
\label{lattice-identity}
\sum_{\bR}e^{-i\bk\cdot\bR}=(2\pi)^d n_0\sum_{\bH}\delta(\bk-\bH)\,.
\ee
By using Eq.~(\ref{eq26}), we rewrite Eq.~(\ref{eq25}) as 
\bea
\label{eq27}
S_\rho(\bk)&=&(m-1)\left[\tilde \phi_s(\bk) -1 \right]\\ &+&m
\sum_{\bR\ne {\bf 0}}e^{-i\bk\cdot\bR} \left[\tilde
\phi_d(\bk;\bR)-1\right]+mS_n(\bk)\,.\nonumber 
\eea 
Note that $S_n(\bk)$ vanishes identically in the first Brillouin
zone, so that it does not contribute to the small $\bk$ expansion of
Eq.~(\ref{eq27}). Therefore, assuming all the moments of the
displacements to be finite, in this region of $k-$space we can
write 
\bea
\label{eq28}
S_\rho(\bk)&=&(m-1)\sum_{l=1}^\infty (-i)^l\frac{\overline{(\bk\cdot
\bw_s)^l}}{l!}\\ &+&m\sum_{\bR\ne {\bf 0}}e^{-i\bk\cdot\bR}
\sum_{l=1}^\infty(-i)^l\frac{\overline{[\bk\cdot
\bw_d(\bR)]^l}}{l!}\,,\nonumber 
\eea
where we have denoted, respectively, $\bw_s$ the relative
displacement of two different particles belonging to the same cloud,
and $\bw_d(\bR)$ the relative displacement of two particles belonging
to two clouds whose centers are separated by the lattice vector $\bR$.
From this formula we can deduce the conditions on the two-displacement
correlations to have a given small $k$ SF for the resulting
particle distribution.

\subsection{Example: lattice of random correlated dipoles}
\label{sec5b}

Let us consider the following example as an application
of Eq.~(\ref{eq28}). Each particle on a perfect lattice is split into
two particles, and the following displacements are applied:
\be
\bu_1(\bR) = + \bfeta (\bR)\,,\qquad
\bu_2(\bR) = - \bfeta (\bR)
\label{eq29}
\ee
where 
${\bf \eta} (\bR)$ is a random vector at each $\bR$ specified by a
lattice translationally invariant correlated stochastic process. The
average over all the realizations of the displacements of 
a function $X( \{ \bfeta (\bR) \} )$ of the displacements may be
written as the functional integral 
\be 
\overline{X}=\prod_{\bR} \int
d^d \bfeta (\bR) {\cal P} \left( \{ \bfeta (\bR) \} \right) X ( \{
\bfeta (\bR) \} )\,. 
\ee
We assume that the joint probability density function of all the
displacements ${\cal P} \left( \{ \bfeta (\bR) \} \right)$ is
invariant under any lattice translation. Moreover it is simple to
show, given our symmetry assumption for the displacements, that we
can take ${\cal P} \left( \{ \bfeta (\bR) \} \right)$ to be invariant
under the change of sign of any {\it individual} $\bfeta (\bR)$.
This ensures that 
$p_d(\bu,\bv;\bx)$ is well-defined as required in our
derivation, i.e., the joint probability for displacements
to two particles at different sites is the same for all couples,
and depends only on their relative separation. With this
assumption it follows that all odd powers $\ell$ in the 
sums in Eq.~(\ref{eq28}) vanish. Calculating the contribution 
at second order in $\bk$ we have
\be
\overline{[\bk\cdot \bw_s]^2}=
4 \overline{[\bk\cdot {\bfeta}]^2}
\ee
and 
\bea
\overline{[\bk\cdot \bw_d(\bR)]^2}
&=& \int d^3 \bfeta({\bf 0}) d^3 \bfeta(\bR) 
p_d[\bfeta({\bf 0}),\bfeta(\bR);\bR] 
\nonumber \\
&& \qquad 
\times [\bk \cdot \left( \bfeta({\bf 0})-\bfeta(\bR) \right) ]^2 \nonumber \\
&=& 2 \overline{[\bk\cdot {\bfeta}]^2}\,.
\eea
The latter results uses the fact that the two point correlation 
function $\overline{\bfeta({\bf 0})\cdot \bfeta(\bR) }=0$, because of the 
assumed inversion symmetry. Using these results in Eq.~(\ref{eq28})
together with the identity Eq.~\ref{lattice-identity}, one finds
that the two contributions cancel in the first Brillouin zone
to give zero. 

At next non-trivial order, fourth order in $\bk$, we find 
\be
\overline{[\bk\cdot \bw_s]^4}= 16 \overline{[\bk\cdot
{\bfeta}]^4} 
\ee 
and 
\bea 
\overline{[\bk\cdot \bw_d(\bR)]^4} &=&
\int d^d \bfeta({\bf 0}) d^d \bfeta(\bR) p_d[\bfeta({\bf 0}),\bfeta(\bR);\bR] \nonumber\\ 
&& \qquad \times [\bk \cdot \left( \bfeta({\bf 0})-\bfeta(\bR) \right)
]^4 \nonumber \\ 
&=& 2 \overline{[\bk\cdot {\bfeta}]^4} + 6
\overline{[\bk\cdot {\bfeta} ({\bf 0})]^2 [\bk\cdot {\bfeta} (\bR)]^2} \,.
\eea 
Using again the identity Eq.~(\ref{lattice-identity}), we obtain
the leading non-trivial contribution to the PS which may be written
\be
\label{dipoles-result-k4}
S_\rho(\bk)\simeq \frac{1}{2} k_\alpha k_\beta k_\gamma k_\delta
\sum_{\bR}e^{-i\bk\cdot\bR} \overline{\eta_\alpha ({\bf 0}) \eta_\beta ({\bf 0})
\eta_\gamma (\bR) \eta_\delta (\bR)}\,, \ee where we adopted the sum
convention on the repeated index $\alpha,\beta,\gamma,\delta$.  The
sum over $\bR$ is just the FT (on the lattice) of a two point
correlation function, the behavior of which as $\bk \rightarrow {\bf 0}$
depends on the nature of these correlations.  The leading small $k$
behaviour of the SF will depend on that of the sum. If the
correlations of the dipoles are short-range, the sum converges to a
positive constant at small $k$, giving a leading behaviour
proportional to $k^4$. If, on the other hand, they are long range
correlated, this sum will diverge as a power-law at small $k$, with an
exponent less than that of the dimension of the space. This will lead
to behaviour of the overall SF proportional to $k^\gamma$ with $4-d <
\gamma \leq 4$. Finally, if the correlations of the dipoles have
themselves superhomogeneous properties\footnote{Note that, as 
the functions $\overline{\eta_\alpha ({\bf 0}) \eta_\beta ({\bf 0}) \eta_\gamma
(\bR)\eta_\delta (\bR)}$ for $\alpha=\beta$ and $\gamma=\delta$ are
non-negative at all $\bR$, their lattice Fourier transforms
can vanish at $\bk={\bf 0}$ only if they are identically zero
for all $\bR$.}, one can obtain such
a behaviour with $\gamma >4$.

\section{Summary and discussion}
\label{sec6}

In this paper we have introduced and analyzed a wide class of non-trivial
stochastic point processes for
which it is possible to write exactly the two-point correlation
function and/or SF.  They are obtained from a given 
``mother'' particle distribution, the SF
of which is assumed known, by
substituting each particle with a cloud of a fixed number $m$
of other particles.  The position of the new particles composing the
clouds is determined by that of the related mother particle plus a
stochastic displacement vector. An important assumption in
all our calculations of the SF is that the stochastic process 
determining the displacements of the particles is independent
of the mother distribution, i.e., the displacements of the 
cloud particles do not depend on the properties, statistical
or otherwise, of the mother distribution to which they are applied.
In practice this means that our SF is defined with respect to
an ensemble average over two independent ensembles: one describing
the realizations of the mother distribution, the other those of
the displacement process.

We have distinguished two cases in which: (i) only the displacements
of different particles belonging to the same cloud may be correlated,
and (ii) the displacements of particle belonging to different
clouds may also be statistically dependent.  In both cases we obtain a
direct generalization of the relations found in \cite{displa} for
$m=1$.  In the first case, once the average over realizations is
taken, the SF of the final particle distribution is related to the SF
of the mother distribution by a local relation in the wave vector
$\bk$, while the second case leads to a more complex relation.  A 
detailed analysis of case (i) led us to find and to discuss, in the 
case of an initial lattice mother distribution, the relations linking
the exponent of the final SF to the number of conserved mass moments
in each cloud: we have seen that such an exponent, if larger or
equal to $4$, tells us directly which are the locally
conserved mass moments in the distribution.  
When we move to case (ii), the presence of cloud-cloud displacement
correlations ``interact'' with the local mass moments conservation in 
determining the small $k$ behavior of the SF of the particle distribution.

One application we have in mind of the results derived here
for this class  of ``stochastically ordered'' point processes 
is in the systematic study of the dynamics of
particle systems driven by long range pair interactions. More 
specifically, in the case of gravity, it is expected (see, e.g., 
\cite{pee80}) that the large scale fluctuations in an infinite 
particle system dominate the dynamics of the gravitational 
clustering for an initial SF with a small $k$ behaviour 
proportional to $k^\gamma$ and $\gamma <4$. This ``hierarchical''
behaviour has been observed numerically for a range of such
$\gamma$, up to a maximal value of $\gamma=2$ (see \cite{sl1}
for a recent discussion, and further references). No
study of the regime of initial conditions $\gamma >4$
has been performed up to now, as no algorithm has been
given in the literature, to our knowledge, which can generate
such an initial condition\footnote{Explicit algorithms 
generating point distribution with $\gamma=4$ have, on 
the other hand, been discussed. See, notably,
\cite{fratzl_etal1999, wandelt, gabrielli_etal}.}. 
It is expected that the gravitational clustering
will be qualitatively different in this case, with 
structures being built up from smaller to larger scales.
Indeed the reason why we expect such a difference can
be understood easily in the context of our constructions
here of such point processes. For $\gamma > 4$ the
fluctuations in mass are so suppressed that gravity
is effectively ``screened'' (at least in the initial
conditions): in a multipole expansion of the mass
far away from a given point in the ``uncorrelated cloud
lattice'' only the leading moment varying from cloud to
cloud will contribute (as the contributions from
the moments which are fixed will cancel out). If this
leading contributing moment is the second moment,
this gives an effectively short-range force 
(decaying as the inverse of the fourth power of
the distance).

Our analysis here has  been limited to the case 
of ``analytic''  exponents for the SF derived from short
tailed displacements probability density functions (PDF).  In
principle one can consider also the case in which such PDFs have a
long power law tail. In this case a singular part of the small $k$
expansion of the related displacement characteristic function arises.
This leads (see \cite{displa, libro}) to a final SF 
characterized by non-analytic
(e.g. fractional) small $k$ exponents. We have seen, however, 
that in the uncorrelated cloud lattice, to attain powers larger
than $\gamma=4$ we must in fact take limited displacements,
and thus we necessarily obtain  ``analytic'' exponents
(and, in fact, a analytic behaviour of the SF at small $k$).
In the case of the correlated cloud lattice, nevertheless, we
have given an example (random correlated dipoles) which shows 
how such non-analytic powers should be attainable by 
including appropriate correlations between the clouds.

We return finally to an important feature of Eq.~(\ref{eq24}) which we
have discussed at some length in Sect.~\ref{sec3}.  This is that the
exponent in the small $k$ scaling behavior of $S_\rho(\bk)$ cannot be
larger than that in $S_n(\bk)$.  As we explained, this can be
understood physically as it means that the replacements of particles
by clouds cannot make the initial particle distribution more
ordered. We underline, however, that this is true given the
assumptions we have made, and specifically assuming that the 
stochastic process generating the
clouds is independent of the mother distribution. In a forthcoming
article with another collaborator \cite{tiling} we will report results
on a related kind of construction of superhomogeneous point processes,
starting from tilings of space with equal volume tiles. In this case
it turns out that one can, in certain circumstances, ascribe a cloud
of particles to each particle of a given point distribution and as a
result increase the exponent $\gamma$. The reason why this becomes
possible is that the displacements of the particles are not applied
independently of the correlation properties of the underlying point
process, as we have assumed here. Indeed in order to increase the
exponent requires that the moments of the clouds are ``tuned''
appropriately to the tile in which the mother point lies. We note also
that, in the case that the initial tiling is taken to be a lattice,
the algorithm described coincides with that given here and similar
results to those given here are recovered.  Details will be reported
in \cite{tiling}.

We thank B. Jancovici, J. Lebowitz and S. Torquato for useful
discussions.   

\appendix

\section{Symmetric cloud lattices}

\label{appa}

In this appendix we study the SF for an uncorrelated cloud process
on a lattice, for the case that each cloud is symmetric with respect to 
its own lattice site. 

The number density of such a particle distribution can be written as:
\begin{equation}
\rho(\bx)=\sum_{\bR}\sum_{j=1}^m\delta(\bx-\bR-\bu^{(\bR)}_j)
\label{eq-rho}
\end{equation}
where $\bR$ is the lattice site and $m$ is the number of particle per
cloud.  Since we consider symmetric clouds, $m$ is even
and can be written as $m=2p$ where $p$ is a positive integer. 
We consider initially a finite lattice with $M$ sites 
(occupying a corresponding finite volume $V$), and then
send $M\to \infty$ at the end of calculations. Taking
the FT of Eq.~(\ref{eq-rho}), we have 
\be
\tilde\rho(\bk,V)=\sum_{\bR}\sum_{j=1}^m
e^{-i\bk\cdot(\bR+\bu^{(\bR)}_j)}\,.
\label{eq-dens-ft}
\ee Now we impose the inversion symmetry for the clouds with respect
to their center of mass (i.e. the lattice site).  This means that for
each particle at $(\bR+\bu)$ there is another particle placed at
$(\bR-\bu)$. Therefore for each cloud we can count with $j$ from $1$
to $p=m/2$ a set of particles which are not the symmetric image one of
each other, and with $p+j$ their respective symmetric images.
Particles in a single cloud with $1\le j\le p$ can be arbitrarily
correlated.  Let us call as above $p_s(\bu,\bv)$ the joint PDF of a
couple of displacements referred to the set of particles with $j=1,
...,p$ in the same cloud and $p(\bu)$ the PDF of a single
displacement. 

Imposing this symmetry of the clouds we can
rewrite Eq.~(\ref{eq-dens-ft}) as 
\be
\tilde\rho(\bk,V)=2\sum_{\bR}e^{-i\bk\cdot\bR}\sum_{j=1}^p
\cos(\bk\cdot\bu^{(\bR)}_j)\,.
\label{eq-dens-ft2}
\ee 
Then taking the squared modulus we obtain
\bea
\label{eq-rho2}
&&|\tilde\rho(\bk,V)|^2\\ &&=4\sum_{\bR,\bR'}e^{-i\bk\cdot(\bR-\bR')}
\sum_{l,j}^{1,p}
\cos(\bk\cdot\bu^{(\bR)}_j)\cos(\bk\cdot\bu^{(\bR')}_l)\nonumber\\
&&=4\left\{\sum_{\bR}\sum_{j=1}^p[\cos(\bk\cdot\bu^{(\bR)}_j)]^2
\right.\nonumber\\ &&+ \sum_{\bR} \sum_{l\ne j}^{1,p}
\cos(\bk\cdot\bu^{(\bR)}_j)\cos(\bk\cdot\bu^{(\bR)}_l) \nonumber\\
&&\left.+\sum_{\bR\ne \bR'}e^{-i\bk\cdot(\bR-\bR')}\sum_{l,j}^{1,p}
\cos(\bk\cdot\bu^{(\bR)}_j)\cos(\bk\cdot\bu^{(\bR')}_l)\right\}
\nonumber
\eea 
We can now take the average over the displacements.  In order to
do this we recall that we are assuming that the displacements 
related to particles belonging to different clouds are uncorrelated.
Consequently the first and the third terms in Eq.~(\ref{eq-rho2}) have
to be averaged over only $p(\bu)$, while the second one as to be
averaged over $p_s(\bu,\bv)$ containing all the two-displacements
correlators in a single cloud. This gives 
\bea
&&\la|\tilde\rho(\bk,V)|^2\ra=4Mp\la[\cos(\bk\cdot\bu)]^2\ra\nonumber\\
&&+4Mp(p-1)\la\cos(\bk\cdot\bu)\cos(\bk\cdot\bv)\ra\nonumber\\
&&+4p^2\sum_{\bR\ne \bR'}e^{-i\bk\cdot(\bR-\bR')}
\la\cos(\bk\cdot\bu)\ra^2 \nonumber\\
&&=2N\la[\cos(\bk\cdot\bu)]^2\ra+N(m-2)\la\cos(\bk\cdot\bu)
\cos(\bk\cdot\bv)\ra\nonumber\\ &&-Nm\la\cos(\bk\cdot\bu)\ra^2+l.t.
\label{eq-rho2-av}
\eea
where $N=Mm$ is the total number of particles and ``$l.t.$'' indicates a 
``lattice term'' which is proportional to the lattice SF
(and which therefore does not contribute around $\bk={\bf 0}$).
In performing the last step of Eq.~(\ref{eq-rho2-av}) we have used the simple
identity
\[\sum_{\bR\ne \bR'}e^{-i\bk\cdot(\bR-\bR')}=\sum_{\bR,\bR'}
e^{-i\bk\cdot(\bR-\bR')}-M\]
which gives rise to the third term of the last step.
Now we use the definition of the SF
\[S_{\rho}(\bk)=\lim_{M\to\infty} \frac{\la|\tilde\rho(\bk,V)|^2\ra}
{N}-(2\pi)^d\rho_0\delta(\bk)\,,\]
which gives
\bea
S_{\rho}(\bk)&=&2\la[\cos(\bk\cdot\bu)]^2\ra+(m-2)\la\cos(\bk\cdot\bu)
\cos(\bk\cdot\bv)\ra\nonumber\\
&-&m\la\cos(\bk\cdot\bu)\ra^2+l.t.
\label{eq-S}
\eea
It is simple to verify that, as expected, $S_\rho({\bf 0})=0$.
We now expand Eq.~(\ref{eq-S})  in power series of $\bk$ 
and study it order by order:
\bea
\label{eq-S-series}
&&S_{\rho}(\bk)=\sum_{a,b}^{0,\infty}\frac{(-1)^{a+b}}{(2a)!(2b)!}
\left[2\la(\bk\cdot\bu)^{2(a+b)}\ra\right.\\
&&\left.+(m-2)\la(\bk\cdot\bu)^{2a}
(\bk\cdot\bv)^{2b}\ra-m\la(\bk\cdot\bu)^{2a}\ra
\la(\bk\cdot\bu)^{2b}\ra\right]\nonumber
\eea
We see that only even powers of $k$ are present. Let us call
${\cal O}_{2n}(k)$ the term of order $k^{2n}$ in the series above.
We see immediately that
\[{\cal O}_{2}(k)=0\]
as all the clouds conserve the center of mass at their lattice site for 
symmetry.
Therefore the first non trivial term is ${\cal O}_{4}(k)$, which after some
manipulation can be written as
\bea
{\cal O}_{4}(k)&=&\frac{1}{2}\left[\la(\bk\cdot\bu)^{4}\ra+(p-1)
\la(\bk\cdot\bu)^{2}(\bk\cdot\bv)^{2}\ra\right.\nonumber\\&-&
\left.p\la(\bk\cdot\bu)^{2}\ra^2\right]\,.\nonumber\\
\eea
The next term is
\bea
{\cal O}_{6}(k)&=&-\frac{1}{12}\left[\la(\bk\cdot\bu)^{6}\ra+(p-1)
\la(\bk\cdot\bu)^{4}(\bk\cdot\bv)^{2}\ra\right.\nonumber\\
&-&\left.
p\la(\bk\cdot\bu)^{4}\ra\la(\bk\cdot\bu)^{2}\ra\right]\,.
\label{eq-O6}
\eea
Both terms are in general non-zero. In order to find the condition
for ${\cal O}_{4}(k)$ to vanish, we rewrite it as
\bea
\label{eq-O4}
&&{\cal O}_{4}(k)=\frac{1}{2}\left\{\la\left[(\bk\cdot\bu)^{2}-
\la(\bk\cdot\bu)^{2}\ra\right]^2\ra\right.\\ &&+(p-1)
\left.\la\left[(\bk\cdot\bu)^{2}-\la(\bk\cdot\bu)^{2}\ra\right]
\left[(\bk\cdot\bv)^{2}-\la(\bk\cdot\bv)^{2}\ra\right]\right\}\ra
\nonumber \eea In any cloud there are $p=m/2$ stochastic displacements
which constitute a closed set of $p$ symmetrically correlated variables
(i.e. the correlation between any pair of these displacement is
constant and there is no correlation with displacements in other
clouds). Since the correlation matrix of the random variables
$\left[(\bk\cdot\bu_j)^2-\la(\bk\cdot\bu)^{2}\ra\right]$, with
$j=1,...,p$ in a single cloud, has to be, as all the correlation
matrix of a set of random variables, positive definite (see above) 
we always have
${\cal O}_{4}(k)\ge 0$.  This can be seen in a more intuitive way by
noticing what follows: \be \la\left\{\sum_{j=1}^p
\left[(\bk\cdot\bu_j)^2-
\la(\bk\cdot\bu_j)^{2}\ra\right]\right\}^2\ra=2p{\cal O}_4(k)
\label{eq-O4-pos}
\ee
which is a variance and consequently manifestly non-negative.
Equation (\ref{eq-O4-pos}) also implies that ${\cal O}_4(k)$ vanishes
if and only if
\[\la\left\{\sum_{j=1}^p \left[(\bk\cdot\bu_j)^2-
\la(\bk\cdot\bu_j)^{2}\ra\right]\right\}^2\ra=0\,,\]
i.e., if with probability 1 we have
\[\sum_{j=1}^p (\bk\cdot\bu_j)^2=p
\la(\bk\cdot\bu)^{2}\ra \,.\] 
This is just the ``conservation law'' of the second 
moment of the mass dispersion
of the clouds in the direction of $\bk$. As the orientation of
$\bk$ is arbitrary, this means that, in order 
to have ${\cal O}_4(k)=0$ identically, the
second moment of the mass of the clouds must be conserved as a tensor
$I_{\mu\nu}=\sum_{j=1}^p u_j^{(\mu)} u_j^{(\nu)}$ with
$\mu,\nu=1,...,d$.

We now analyze ${\cal O}_6(k)$ which is given by Eq.~(\ref{eq-O6}).
First of all we note that 
\be
\la\left[\sum_{j=1}^p (\bk\cdot\bu_j)^4\right]\sum_{l=1}^p
\left[(\bk\cdot\bu_l)^2-\la(\bk\cdot\bu_l)^{2}\ra\right]\ra=-12p{\cal
O}_6(k)\,.
\label{eq-O6-id}
\ee
But, as seen above, if ${\cal O}_4(k)=0$ the second sum in 
Eq.~(\ref{eq-O6-id}) vanishes identically, and therefore we can
conclude that when ${\cal O}_4(k)=0$ identically also ${\cal O}_6(k)=0$
automatically, and the dominant term in $S_\rho(k)$ becomes ${\cal O}_8(k)$.

One can continue this analysis further and show, after
some more involved algebra, that the dominant term in the small $k$
expansion of the SF is of order $k^{4n}$ with $n$ integer, and next
order terms are proportional to $k^{4n+2q}$ with  $q$ again an integer.
The exponent $n$ depends on the order to which the moments
of the mass dispersion of the clouds are conserved.

\end{document}